\documentclass{elsart}   

\newcommand{\vct}[1]{\mbox{\boldmath{$#1$}}}
\newcommand{\pfermi}{p_{\mathrm{F}}}
\newcommand{\efermi}{E_{\mathrm{F}}}
\newcommand{\vfermi}{v_{\mathrm{F}}}
\newcommand{\nfermi}{N_{\mathrm{F}}}
\newcommand{\fsigma}{F_{\mathrm{s}}}
\newcommand{\fomega}{F_{\mathrm{v}}}
\newcommand{\nbar}{\bar{\mathrm{N}}}
\newcommand{\msigma}{m_{\mathrm{s}}}
\newcommand{\momega}{m_{\mathrm{v}}}
\newcommand{\gsigma}{g_{\mathrm{s}}}
\newcommand{\gomega}{g_{\mathrm{v}}}
\newcommand{\pih}{\Pi_{\mathrm{H}}}
\newcommand{\pisigma}{\Pi_{\mathrm{s}}}
\newcommand{\piomega}{\Pi_{\mathrm{v}}}
\newcommand{\proh}{G_{\mathrm{H}}}
\newcommand{\prohole}{G_{\mathrm{h}}}
\newcommand{\psih}{\psi_{\mathrm{H}}}
\newcommand{\proden}{G_{\mathrm{D}}}
\newcommand{\profyn}{G_{\mathrm{F}}}
\newcommand{\pronbar}{G_{\bar{\mathrm{N}}}}
\newcommand{\chiomega}{\chi_{\mathrm{v}}}
\newcommand{\chisigma}{\chi_{\mathrm{s}}}
\newcommand{\aomega}{a_{\mathrm{v}}}
\newcommand{\asigma}{a_{\mathrm{s}}}

\begin{document}
\begin{frontmatter}
\title{Effects of the Dirac Sea on the Giant Monopole States}

\author{Haruki Kurasawa}
\address{Department of Physics, Faculty of Science, Chiba
University,\\ Chiba 263-8522, Japan}

\author{Toshio Suzuki}
\address{Department of Applied Physics, Fukui University, Fukui
910-8507, Japan}
\address{RIKEN, 2-1 Hirosawa, Wako-shi, Saitama 351-0198, Japan}

\begin{abstract}
Effects of the Dirac sea on the excitation energy
of the giant monopole states are investigated in an analytic way
within the $\sigma-\omega$ model. The excitation energy is determined
by the relativistic Landau-Migdal parameters, $F_0$ and $F_1$.
Their analytic expressions are derived in  the relativistic random
phase approximation(RRPA) without the Dirac sea, with the Pauli
blocking terms and with the full Dirac sea. It is shown that in
the RRPA based on the mean field approximation the Pauli blocking
terms should be included in the configuration space, according
to the relativistic Landau theory. In the renormalized RRPA, the
incompressibility coefficient becomes negative, if N$\nbar$
excitations are neglected.
\end{abstract}

\end{frontmatter}

The relativistic mean field approximation(RMFA) neglects the Dirac
sea in the description of the nuclear ground state.
Recently, however, it has been numerically shown that 
in the relativistic random phase approximation(RRPA) built on the
RMFA, the monopole states cannot be  well described without the
Pauli blocking terms which express transitions between the Dirac
sea and the occupied Fermi sea \cite{giai}.
If the blocking terms are neglected, the excitation energies of
the monopole states in the RRPA are much lower than those in the
time-dependent relativistic mean field approximation \cite{ring}.

The purpose of the present paper is to show in an analytic way
the role of the Dirac sea in the excitation energy of the monopole
states. We will discuss the monopole states of nuclear matter in terms
of the Landau-Migdal parameters using the $\sigma-\omega$ model.
First, we will show that in the RRPA based on the RMFA, one should
take into account the Pauli blocking terms in the configuration
space. Second, the real effects of the Dirac sea will be discussed
in the renormalized RRPA. It will be shown that N$\bar{\mathrm{N}}$
states yield essential effects on the excitation energy through
the Landau-Migdal parameter, $F_0$.

The Landau-Migdal parameters, $F_0$ and $F_1$, are obtained by
the second derivative of the total energy density with respect
to the quasiparticle distribution. In the RMFA, they are given
by \cite{mat,bak} 
\begin{equation}
 F_0 = \fomega - \frac{1 - \vfermi^2}{1 + \asigma\fsigma}
 \fsigma, \quad F_1 = -\frac{\vfermi^2\fomega}
 {1 + \frac{1}{3}\vfermi^2\fomega},\label{par}
\end{equation}
where we have defined
\begin{equation}
\fsigma = \nfermi\left(\frac{\gsigma}{\msigma}\right)^2,\quad
\fomega = \nfermi\left(\frac{\gomega}{\momega}\right)^2,
\end{equation}
\begin{equation}
\nfermi = \frac{2\pfermi \efermi}{\pi^2},\quad
\vfermi = \frac{\pfermi}{\efermi},\quad
\efermi = ( \pfermi^2 + M^{*2})^{1/2}. 
\end{equation}
In the above equations, $\gsigma$ and $\gomega$ stand for the
Yukawa coupling constants, $\msigma$ and $\momega$ the
masses of the $\sigma$- and $\omega$-meson, respectively, and 
$\pfermi$ and $M^*$ denote the Fermi momentum and the effective
nucleon mass. $ \nfermi$ and $\vfermi$ represent the relativistic
density of states at the Fermi surface and the relativistic Fermi
velocity.
The factor, $\asigma$, in $F_0$ of Eq.(\ref{par}) will play an
essential role  in later discussions, which is given by
\begin{equation}
 \asigma\fsigma = \frac{4}{(2\pi)^3}
 \left( \frac{\gsigma}{\msigma} \right)^2
 \int d^3p\,\frac{\vct{p}^2}{E_p^3}\theta_p,\quad
 E_p = ( \vct{p}^2 + M^{*2})^{1/2},\label{pau} 
\end{equation}
where $\theta_p$ denotes the step function, $\theta(\pfermi -
|\vct{p}|)$.
 
In the relativistic model, the excitation energy of the monopole
states is expressed as \cite{nks},
\begin{equation}
E_{\mathrm{M}}
= \left(\frac{K}{\epsilon_{\mathrm{F}}\langle r^2\rangle}\right)^{1/2},
\label{em}
\end{equation}
where $\epsilon_{\mathrm{F}}$ denotes the Fermi energy and $\langle
r^2\rangle$ the root mean square radius of the nucleus.
The incompressibility coefficient, $K$, is expressed in terms of
the above relativistic Landau-Migdal parameters, 
\begin{equation}
K = \frac{3\pfermi^2}{\epsilon_{\mathrm{F}}}
    \frac{1 + F_0}{1 + \frac{1}{3}F_1}.\label{com}
\end{equation}
Since $\pfermi$ is determined by the nucleon density, and
$\epsilon_{\mathrm{F}}$ is related to the nucleon binding energy,
$E_{\mathrm{B}}$, and the free nucleon mass, $M$,
\begin{equation}
\epsilon_{\mathrm{F}} = E_{\mathrm{B}} + M,\label{fe}
\end{equation}
the excitation energy of the monopole state is a function of $F_0$
and $F_1$.

In order to see the effects of the Pauli blocking terms on the
monopole states, we derive the Landau-Migdal parameters according
to the RRPA. We calculate the longitudinal RRPA correlation functions
with and without the Pauli blocking terms. By comparing them with
the correlation function of the Landau theory\cite{lan}, we will
obtain the expressions of the Landau-Migdal parameters in each
approximation.

When following our previous papers \cite{bak,ks}, 
the mean field correlation function, $\pih$ \cite{bak,ks},
is  given by the Fourier transform of the single-particle Green
function, $\proh$, 
\begin{equation}
\pih(A,B;k) = -\frac{1}{2\pi i}\int d^4p\,
 {\mathrm{Tr}}[\Gamma_A\proh(p+k)\Gamma_B\proh(p)],\label{cor}
\end{equation}
where $k$ denotes the four-momentum, $(k_0,\vct{k})$, and
$A$ and $B$ the Fourier transform of the external field
expressed with the mean field, $\psih(\vct{x})$,
\begin{equation}
 A(\vct{k}) = \int d^3x \exp(i\vct{k}\cdot\vct{x})
  \bar{\psi}_{\mathrm{H}}(\vct{x})\Gamma_A\psih(\vct{x}),
  \label{ex}
\end{equation}
$\Gamma_A$ being some 4$\times$4 matrices. 
The sum of the ring diagrams in the RRPA for the $\sigma-\omega$
model is described as \cite{bak,ks},
\begin{equation}
 \delta\Pi_{\mathrm{RPA}}(A, B;k)
 = \frac{\chisigma\tilde{\chi}_{\mathrm{v}}}
 { {\mathrm{det}} U_{\mathrm{L}}}
 \pih(A,\Lambda^a;k)(U_{\mathrm{L}})_{ab}
 \pih(\Lambda^b, B;k),\label{ring}
\end{equation}
where the contraction should be carried out with respect to the
superfix and suffix, $a, b = -1, 0$, and $\Lambda_{-1}$ and $\Lambda_0$
are given by Eq.(\ref{ex}) with $\Gamma_A = 1$ and $\gamma_0$,
respectively.
Moreover, $\chisigma$ and $\tilde{\chi}_{\mathrm{v}}$ represent,
\begin{equation}
\chisigma = \frac{1}{(2\pi)^3}
 \frac{\gsigma^2}{\msigma^2 - k^2},\quad 
 \tilde{\chi}_{\mathrm{v}} = \frac{1}{(2\pi)^3}
 \frac{\gomega^2}{\momega^2 - k^2}\frac{k^2}{\vct{k}^2}.
\end{equation}
The explicit form of the 2$\times$2 matrix, $U_{\mathrm{L}}$, in
Eq.(\ref{ring})
depends on whether or not the Pauli blocking terms are included
in the mean field correlation functions as discussed below.

The Green function, $\proh$, is given by the sum of those for
a single-particle, hole and antinucleon,
\begin{equation}
\proh = G_{\mathrm{p}}( 1 - \theta_p) + \prohole\theta_p
 + \pronbar.
\end{equation}
It is rewritten as a sum of the density-dependent and the Feynman
part \cite{ks},
\begin{eqnarray}
 \proh = \proden + \profyn,\quad
 \proden = \theta_p(\prohole - G_{\mathrm{p}}),\quad
 \profyn = G_{\mathrm{p}} + \pronbar.
\end{eqnarray}
Hence, $\pih$ is composed of the four terms like
$\proden\proden$, $\proden\profyn$, $\profyn\proden$ and
$\profyn\profyn$ \cite{ks}. In the RRPA based on the RMFA, the
$\profyn\profyn$ term is neglected, which is divergent, while
in the previous calculations \cite{bak}, the
$\proden\profyn$ and
$\profyn\proden$ terms, which contain the Pauli blocking
N$\bar{\mathrm{N}}$ excitations like
$G_{\mathrm{p}}\theta_p\pronbar$,
have been kept. Then we have obtained 
\begin{equation}
U_{\mathrm{L}} = \left(
\begin{array}{cc}
\chisigma( 1 - \tilde{\chi}_{\mathrm{v}}\piomega ) &
\chisigma\tilde{\chi}_{\mathrm{v}}\Pi_{\mathrm{sv}}\\
\chisigma\tilde{\chi}_{\mathrm{v}}\Pi_{\mathrm{sv}} &
\tilde{\chi}_{\mathrm{v}}( 1 - \chisigma\pisigma )
\end{array}\right),
\label{dfu}
\end{equation}
where the mean field correlation functions are defined as
\begin{equation}
\pisigma = \pih(\Lambda_{-1}, \Lambda_{-1}; k),\quad
\piomega = \pih(\Lambda_0, \Lambda_0;k),
\end{equation}
\[
\Pi_{\mathrm{sv}} = \pih(\Lambda_{-1}, \Lambda_0;k) =  \Pi_{\rm
H}(\Lambda_0, \Lambda_{-1};k).
\]
The Landau prescription of the correlation functions is obtained
at the limit $k \rightarrow 0$. In this limit we have 
\begin{equation}
\pisigma = (2\pi)^3\nfermi\{( 1 - \vfermi^2 )\Phi(x)
 - \asigma \},\label{s}
\end{equation}
\begin{equation}
\piomega = (2\pi)^3\nfermi\Phi(x), \ \ \ 
\Pi_{\mathrm{sv}} = (2\pi)^3\nfermi( 1 - \vfermi^2 )^{1/2}\Phi(x),
\label{v}
\end{equation}
where $\Phi(x)$ stands for the Lindhard function with
$x = k_0/(|\vct{k}|\vfermi)$ \cite{bak}.
Using these equations, we obtain the generalized dielectric function
from the factor of Eq.(\ref{ring}) as
\begin{eqnarray}
\frac{1}{\chisigma\tilde{\chi}_v}{\mathrm{det}}U_{\mathrm{L}}
 = ( 1 + \asigma \fsigma )
 \left\{ 1 + \left( \fomega
 - \frac{1-v^2_{\mathrm{F}}}{1+\asigma\fsigma}
 \fsigma  - \vfermi^2\fomega x^2 \right)\Phi(x)
\right\}.\label{df}
\end{eqnarray}
In the Landau theory, Eq.(\ref{df}) should be written as \cite{lan}
\begin{equation}
\frac{1}{\chisigma\tilde{\chi}_{\mathrm{v}}}
 {\mathrm{det}}U_{\mathrm{L}} =
c\left\{ 1 + \left( F_0 + \frac{F_1}{1 + \frac{1}{3}F_1}x^2\right)
  \Phi(x)\right\}.\label{lan}
\end{equation}
By comparing Eq.(\ref{df}) with (\ref{lan}), we obtain the
Landau-Migdal parameters which are the same as in Eq.(\ref{par}).

Next we investigate the role of the Pauli blocking terms in the
Landau-Migdal parameters. We calculate the mean field correlation
functions neglecting the Pauli blocking terms and taking the only
particle-hole states. The calculation of the mean field correlation
functions is a little different from the one in taking the Pauli
blocking terms, since the correlation functions are not Lorentz
covariant, and the continuity equation is provided in a different
way,
\begin{equation}
k_\mu\pih(A, \Lambda^\mu;k) = \langle [ \Lambda_0(\vct{k}),
A^\dagger(\vct{k}) ] \rangle,\label{con}
\end{equation}
where $\Lambda^\mu$ is given by replacing $\Gamma_A$ with
$\gamma^\mu$ in Eq.(\ref{ex}), and the r.h.s. is related to the
expectation value of the ground state as,
\begin{equation}
\langle \ |[ \Lambda_0(\vct{k}), A^\dagger(\vct{k'}) ]|\ \rangle
= \delta( \vct{k} - \vct{k'} )
\langle [ \Lambda_0(\vct{k}), A^\dagger(\vct{k'})] \rangle.
\end{equation}
In including the Pauli blocking terms, the r.h.s. of Eq.(\ref{con})
is always vanished, but in neglecting them, it is not for
$\Lambda_{1,2,3}$.
Hence, the relationship between the correlation functions due to
the time- and the longitudinal component of the $\omega$-meson
is written in the frame, $\vct{k} = ( |\vct{k}|, 0, 0 )$, as
\begin{equation}
\pih(\Lambda_1, \Lambda_1;k)
 = \frac{k_0^2}{|\vct{k}|^2}\pih(\Lambda_0, \Lambda_0;k)
 - \aomega,\label{tl}
\end{equation}
where the additional term, $\aomega$, comes from the r.h.s. of
Eq.(\ref{con}),
\begin{equation}
 \aomega
  = \langle [ \Lambda_0(\vct{k}), \Lambda^{1\dagger}(\vct{k})]\rangle
  /|\vct{k}|.\label{acon}
\end{equation}
Because of this fact, the back flow effects due to the longitudinal
$\omega$-meson exchange on $\delta\Pi_{\mathrm{RPA}}$ are not simply
normalized as $\tilde{\chi}_{\mathrm{v}}$, and $U_{\mathrm{L}}$ in
this case depends on $\aomega$,
\begin{equation}
U_{\mathrm{L}}
 = \left(\begin{array}{cc}
  \chisigma( 1 + \aomega\chiomega
- \hat{\chi}_{\mathrm{v}}\piomega ) &
  \chisigma\hat{\chi}_{\mathrm{v}}
  \Pi_{\mathrm{sv}}\\ \chisigma\hat{\chi}_{\mathrm{v}}
  \Pi_{\mathrm{sv}} &
 \hat{\chi}_{\mathrm{v}}( 1 - \chisigma\pisigma )
 \end{array}
  \right)\label{ddu},
\end{equation}
where we have defined 
\begin{equation}
\chiomega = \frac{1}{(2\pi)^3}
 \left( \frac{\gomega}{\momega}\right)^2,\quad 
\hat{\chi}_{\mathrm{v}} = \tilde{\chi}_{\mathrm{v}}
- \aomega\chiomega^2.
\end{equation}

At the limit, $k \rightarrow 0$, $\chiomega\aomega$ becomes
to be
\begin{eqnarray}
\chiomega\aomega(k \rightarrow 0)
= -\,4\left(\frac{\gomega}{\msigma}\right)^2
\int \frac{d^3p}{(2\pi)^3}\,\frac{2\vct{p}^2/3 + M^{*2}}{E_p^3}
\theta_p
= -\frac{1}{3}\vfermi^2\fomega.
\end{eqnarray}
Moreover, $\pisigma$ of the present case has not the term,
$\asigma$, in Eq.(\ref{s}), while $\piomega$ and
$\Pi_{\mathrm{sv}}$ are the same as in Eq.(\ref{v}).
As a result the generalized dielectric function in this case is
given by
\begin{eqnarray}
& &\frac{1}{\chisigma\hat{\chi}_{\mathrm{v}}}
 {\mathrm{det}}U_{\mathrm{L}} \nonumber \\
 \noalign{\vskip4pt}
&=& \left( 1 - \frac{\vfermi^2}{3}\fomega\right)
 \left\{ 1 + \left( \fomega -  \fsigma( 1 - \vfermi^2)
 -  \frac{\vfermi^2\fomega}
 {1 - \frac{1}{3}\vfermi^2\fomega}x^2\right)\Phi(x)\right\}.
 \label{dd}
\end{eqnarray}
Finally comparison of the above equation with Eq.(\ref{lan}) provides
us with the Landau-Migdal parameters in neglecting the Pauli blocking
terms,
\begin{equation}
F_0 = \fomega - ( 1 - \vfermi^2)\fsigma, \quad
F_1 = -\vfermi^2\fomega.\label{dpar}
\end{equation}

The difference between Eqs.(\ref{par}) and (\ref{dpar}) is very
clear. $F_0$ and $F_1$ in Eq.(\ref{dpar}) have no denominator.
In order to obtain the correct expressions of $F_0$ and $F_1$ within
the RMFA, thus we need to include the Pauli blocking terms in the
configuration space of the RRPA.

In the Landau prescription, the denominators in Eq.(\ref{par})
come from the self-consistent derivative of the effective mass
and the baryon current with respect to the quasi-particle
distribution, $n_i$ \cite{mat,nks}. 
As to the effective mass in the RMFA,
\begin{equation}
 M^\ast=M-\left(\frac{\gsigma}{\msigma}\right)^2
       \frac{1}{V}\sum_in_i\frac{M^\ast}{E_{p_i}},
\end{equation}
we have
\begin{equation}
 \frac{\partial M^\ast}{\partial n_j}=
-\,\frac{1}{V}\left(\frac{\gsigma}{\msigma}\right)^2
\frac{M^\ast}{E_{p_j}}
-\left(\frac{\gsigma}{\msigma}\right)^2
       \frac{1}{V}\sum_in_i\frac{\vct{p}_i^2}{E^3_{p_i}}
       \frac{\partial M^\ast}{\partial n_j},\label{mass}
\end{equation}
$V$ being the nuclear volume. The coefficient of
$\partial M^\ast/\partial n_j$ in the r.h.s. yields
$\asigma\fsigma$ in the denominator of $F_0$, as seen
in Eq.(\ref{pau}).
In the Green function formalism, the effective mass of the RMFA
is written as
\begin{equation}
M^\ast=M+i\left(\frac{\gsigma}{\msigma}\right)^2
       \int\!\frac{d^4p}{(2\pi)^4}\,{\mathrm{Tr}}\,\proden(p).
\end{equation}
Hence the coefficient of $\partial M^\ast/\partial n_j$ in
Eq.(\ref{mass}) is given by 
\begin{equation}
\asigma\fsigma=
-i\left(\frac{\gsigma}{\msigma}\right)^2
       \int\!\frac{d^4p}{(2\pi)^4}\,{\mathrm{Tr}}\left(
       \frac{\partial \proden(p)}{\partial M^\ast}\right).
\end{equation}
On the other hand, we have shown that
\begin{equation}
\asigma\fsigma = - \frac{1}{(2\pi)^3}
 \left(\frac{\gsigma}{\msigma}\right)^2
 \Pi_{\mathrm{Pauli}}(k=0),
\end{equation}
where $\Pi_{\mathrm{Pauli}}$ represents the Pauli blocking terms
in Eq.(\ref{cor}) for $\Gamma_A = \Gamma_B = 1$.
Thus it is seen that 
in the Landau prescription of the RMFA, the derivative of
$\proden$ includes implicitly the Pauli blocking terms.
Indeed, we can prove that
\begin{equation}
\frac{\partial \proden(p)}{\partial M^\ast}
 = \proden\pronbar + \pronbar\proden
 + \proden\frac{M^*}{p_0}\frac{\partial}{\partial p_0}.
\end{equation}
If we integrate the r.h.s. over $p_0$, the Pauli blocking terms
only remain.

The same discussion is possible for the denominator of $F_1$. The
self-consistent derivative of the current, $\vct{j}$, as to the
quasi-particle distribution provides \cite{mat},
\begin{equation}
\frac{\partial \vct{j}}{\partial n_j}
 = \frac{1}{V}\frac{\vct{p}_j}{E_{p_j}}
- \frac{1}{3}\vfermi^2\fomega
\frac{\partial \vct{j}}{\partial n_j}.\label{mfc}
\end{equation}
The coefficient of the second term yields the denominator of $F_1$.
Using the Green function, the current of the RMFA is written as,
\begin{equation}
\vct{\Sigma} = \left(\frac{\gomega}{\momega}\right)^2\vct{j}
= -\,i \left(\frac{\gomega}{\momega}\right)^2\int
\frac{d^4p}{(2\pi)^4}\,{\mathrm{Tr}}(\vct{\gamma}\proden(p')),\quad
\vct{p}' = \vct{p} - \vct{\Sigma}.
\end{equation}
Hence, the coefficient of the second term in Eq.(\ref{mfc}) is
given by
\begin{equation}
-\,\frac{1}{3}\vfermi^2\fomega\delta_{ij}
= -\,i\left(\frac{\gomega}{\momega}\right)^2
\int\frac{d^4p}{(2\pi)^4}\,{\mathrm{Tr}}
\left.
 \left( \gamma^j \frac{\partial \proden(p')}{\partial {\Sigma}^i}
 \right)\right|_{\vct{\Sigma}=0}.
\end{equation}
We can show that the above derivative of $\proden$ required in
the RMFA is expressed by using the Pauli blocking terms as
\begin{equation}
\frac{\partial \proden}{\partial {\Sigma}^i}
 = \proden\gamma_i\pronbar
 + \pronbar\gamma_i\proden
 - \proden\frac{p'^i}{E_{p'}}\frac{\partial}{\partial p_0}.
\end{equation}

Let us explore the effects of the Pauli blocking terms in more
detail.
The Pauli blocking terms reduce always the contribution of the
$\sigma$-meson to $F_0$ through the factor $\asigma$ in
Eq.(\ref{par}), since $\asigma$ is positive, as seen in
Eq.(\ref{pau}). On the other hand, the contribution of the
$\omega$-meson to $F_0$ is not affected.
Therefore, the value of $F_0$ becomes always smaller, and the
incompressibility, $K$, is reduced according to Eq.(\ref{com}),
when the Pauli blocking terms are neglected. 
On the contrary, the absolute value of $F_1$ in Eq.(\ref{dpar}),
which has no denominator, becomes always larger, compared with
the correct one in Eq.(\ref{par}), so that $K$ is enhanced in
neglecting the Pauli blocking terms.

We calculate the values of $F_0$, $F_1$ and $K$ using the following
parameters \cite{hs} as an example,
\[
M = 939,\quad \msigma = 520,\quad \momega = 783
 \ \ ({\mathrm{MeV}}),
\]
\begin{equation}
\gsigma^2 = 109.626,\quad \gomega^2 = 190.431,
\end{equation}
which reproduce the nucleon binding energy, $E_{\mathrm{B}} = -15.75$
MeV at $\pfermi = 1.30$ fm$^{-1}$. In this case, we have
$M^* = 0.541M$, $\vfermi = 0.451$ and $\asigma = 9.07\times10^{-3}$.
These values provide us with
\begin{equation}
F_0 = 0.569, \quad F_1 = -1.151, \quad K = 544 \  {\mathrm{MeV}}
\end{equation}
in taking the Pauli blocking terms, and
\begin{equation}
F_0 = -0.368, \quad F_1 = -1.866, \quad  K = 357 \ {\mathrm{MeV}}
\end{equation}
in neglecting the Pauli blocking terms. Thus both Landau-Migdal
parameters are strongly affected by the Pauli blocking terms, and,
in particular, $F_0$ changes its sign. Consequently, the value
of $K$ is fairly reduced in neglecting the Pauli blocking terms.
In $^{208}$Pb, the reduction amounts to about 2.7 MeV for the present
parameters. This fact may be observed in ref. \cite{giai} by numerical
calculations. 

Now real effects of the Dirac sea should be explored with 
the fully renormalized RRPA, where the $\profyn\profyn$ term
is also calculated in the Hartree correlation functions. Then,
the Pauli exclusion principle in both the Fermi sea and the Dirac
sea is correctly taken into account. Such calculations based on
the renormalized Hartree approximation(RHA) have been done by the
present authors in ref. \cite{dks}. For complete discussions,  we
quote those results here. The Landau-Migdal parameters in this
case are given by
\begin{equation}
F_0 = F_\omega - \alpha_{\mathrm{ren}} \fsigma,\quad
F_1 = -\frac{\vfermi^2F_\omega}{1 + \frac{1}{3}\vfermi^2F_\omega},
\label{rpar}
\end{equation}
where we have used the abbreviations:
\begin{equation}
F_\omega = \nfermi\left(\frac{\gomega}{m_0}\right)^2,\quad
 \alpha_{\mathrm{ren}}
 = \frac{1 - \vfermi^2}{1 + \asigma\fsigma + a_{\mathrm{D}}}.
 \label{dir}
\end{equation}
Formally the above equation is similar to Eq.(\ref{par}), but the
mass of the $\omega$-meson is replaced by the bare mass, $m_0$,
in $F_\omega$ and the Dirac sea yields an additional effect,
$a_{\mathrm{D}}$ in $\alpha_{\mathrm{ren}}$ \cite{dks}.

As to $F_1$, essentially there is no additional effects from the
Dirac sea. 
The renormalized correlation function from the $\profyn\profyn$
term due to the $\omega$-meson exchange disappears at the limit
$k \rightarrow 0$ and has no contribution to $F_1$. Replacement
of $\momega$ by $m_0$ comes from the fact that the $\omega$-meson
propagator is written in terms of the bare mass\cite{dks}. The
value of $F_1$, however, depends on those of the Yukawa coupling
constants used in the RHA.
In order to reproduce the nucleon binding energy and the Fermi
momentum mentioned before\cite{dks}, the RHA requires $\gsigma^2
= 66.117$ and $\gomega^2 = 79.927$. These values give $M^* =
0.7306M$ and $m_0 = 691.171$ MeV, so that we obtain $F_1 = -0.620$.

On the other hand, $F_0$ is strongly affected by the Dirac sea
through $a_{\mathrm{D}}$ in Eq.(\ref{dir}).
In using $\vfermi$ = 0.3502 of the present RHA,
its value is much larger than that of $\asigma\fsigma$,
\begin{equation}
a_{\mathrm{D}} = 0.405, \quad \asigma\fsigma = 0.0296.
\end{equation}
This Dirac sea effects reduce strongly the contribution of the
$\sigma$-meson to $F_0$ through $\alpha_{\mathrm{ren}}$, and we
have
\begin{equation}
F_0 = 0.676.
\end{equation} 
If $a_{\mathrm{D}}$ were neglected, then the value of $F_0$ would
be $ -1.56$, which means $K < 0$. This fact reflects that
$\bar{{\mathrm{N}}}$-degrees of freedom play an important
role to stabilize the nucleus in the RHA.

Finally we give two comments. First, 
since the restoring force of the giant quadrupole states comes
mainly from the distortion of the kinetic energy density, its
excitation energy depends on $F_1$\cite{nks},
\begin{equation}
E_{\mathrm{Q}}
= \left( \frac{6\pfermi^2}{5\epsilon_{\mathrm{F}}^2\langle
r^2 \rangle}\frac{1}{1 + \frac{1}{3}F_1} \right)^{1/2}.
\label{eq}
\end{equation}
Thus, the Pauli blocking terms affect also the excitation energy
of the quadru\-pole states in the RRPA based on th RMFA. Moreover,
the Pauli blocking terms should be taken into account 
in the description of the center of mass motion, which requires
the correct $F_1$ \cite{nks}. This fact has been observed in
arguments on the spurious state of RRPA by Dawson and
Furnstahl \cite{fur}.
In the same way the Pauli blocking terms  are necessary for
discussions of the nuclear current or magnetic moments \cite{pv}.
The isovector dipole states depend on $F_1$ and $F'_1$, which
also require the Pauli blocking terms.
The detail will be published in a separate paper.

Second, 
we note that  Eqs.(\ref{em}), (\ref{com}) and (\ref{eq}) are formally
the same as those in nonrelativistic models, except for
$\epsilon_{\mathrm{F}}$
in their denominators \cite{nks}. In nonrelativistic models, it
is replaced by the nucleon mass, $M$. However, they are related
to each other as Eq.(\ref{fe}) through the nucleon binding energy
which is negligible compared with the nucleon mass.
Thus the relativistic correction to the excitation energies of
the monopole and quadrupole states is less than 1\% for
$E_{\mathrm{B}} = -15.75$ MeV, if the values of the Landau-Migdal
parameters in relativistic models are the same as in
nonrelativistic ones.

In conclusion, in the relativistic random phase approximation(RRPA)
based on the relativistic mean field approximation, the Pauli
blocking terms should be taken into account for consistent
descriptions of the Landau-Migdal parameters. The real effects
of the Dirac sea on the excitation energy of the monopole states
are studied in the RRPA built on the renormalized Hartree
approximation. Then the nucleon-antinucleon states affect strongly
the excitation energy of the monopole states through the
Landau-Migdal parameter, $F_0$.
The incompressibility coefficient becomes negative,
if antinucleon-degrees
of freedom are neglected. The Landau-Migdal parameter, $F_1$, is
not affected formally by the renormalization.

\begin{ack}
The authors would like to thank Professor
Nguyen Van Giai and Professor P. Ring for useful discussions.
\end{ack}

\end{document}